\documentclass[12pt,twocolumn]{iopart}
\usepackage[pdftex]{graphicx}
\begin{document}
\title[]
{Comparison of porous silicon prepared using metal - induced etching (MIE) and laser - induced etching (LIE)}
\author{Shailendra K Saxena$^1$,Vivek Kumar$^2$, Hari M. Rai$^1$, Gayatri Sahu$^1$,Ravi K. Late$^1$, Kapil Saxena$^3$, A. K. Shukla$^3$, Pankaj R Sagdeo$^{1,4}$ and Rajesh Kumar$^{1,4}$ \footnote{Corresponding Author: rajeshkumar@iiti.ac.in}}
\address{$^1$Material Research Laboratory, Discipline of Physics, School of Basic Sciences, Indian Institute of Technology Indore, Madhya Pradesh-452017, India}
\address{$^2$ Biophysics \& Nanoscience Centre, DEB-CNISM, Universita della Tuscia, I-01100 Viterbo, Italy}
\address{$^3$Department of Physics, Indian Institute of Technology Delhi, Hauz Khas, New Delhi 110016, India}
\address{$^4$Material Science and Engineering Group, Indian Institute of Technology Indore, Madhya Pradesh-452017, India}
\begin{abstract}
Porous silicon (p-Si), prepared by two routes (metal induced etching (MIE) and laser induced etching (LIE)) have been studied by comparing the surface morphologies. A uniformly distributed smaller (submicron sized) pores are formed when MIE technique is used because the pore formation is driven by uniformly distributed metal (silver in present case) nanoparticles, deposited prior to the porosification step. Whereas in p-Si samples prepared by LIE technique,  wider  pores with some variation in pore size as compared to MIE technique is observed because a laser having gaussian profile of intensity is used for porosification.  Uniformly distribute well-aligned Si nanowires are observed in samples prepared by MIE method as seen using cross-sectional SEM imaging, which shows a single photoluminescence (PL) peak at 1.96 eV corresponding to red emission at room temperature. The single PL peak confirms the presence of uniform sized nanowires in MIE samples. These vertically aligned Si nanowires can be used for field emission application.  

\end{abstract}

\maketitle

\section{Introduction}

\ \ \ \ Attention received by porous silicon (p-Si) from scientists is known from more than couple of decades since the discovery of its light emission properties [1,2]. p-Si is still being studied due to its applications in various fields in addition to optoelectronic devices [1-7]. Many reports shows applications of p-Si in advanced sensing[8-11], in making electrodes for batteries [12,13], in carrying out mass spectroscopy [14] in carrying out mass spectroscopy [15] and in field effect transistors [16, 17]. Different methods to fabricate p-Si \& it’s various applications/properties have been studied in depth and very good reports are available [18, 19]. Various techniques used for the formation of p-Si include strain etching [20, 21] of Si wafer in a solution of HNO $_3$ \& HF acid, electrochemical etching [22, 23] where an electrical bias is used for porosification of Si wafer in etching solution of HF acid. A simpler method of photochemical etching also known as laser induced etching (LIE) was used by Choy and Cheah [24] and later used by many others [25 - 29].  LIE is mainly used for preparing p-Si in a localized area on a Si wafer without using an electrical bias. A combination of two techniques (LIE and electrochemical etching) has also been reported [30, 31] for fabrication of p-Si. All these methods have certain advantages and disadvantages over each other and are used depending on the convenience \& suitability. LIE method and stain etching method does not require any bias for making p-Si whereas a bias is needed while using electrochemical etching.  Stain etching is the easiest method to use but is a very slow process and does not give a `reproducible \& reliable’ result [32].

In addition to the above-mentioned  techniques, method of metal assisted chemical etching or metal induced etching (MIE)  [32–38] is also used for preparing p-Si at faster rates as compared to stain etching. The MIE technique, in contrast to stain etching technique, gives good reproducibility in p-Si formation without using any bias or laser radiation. Moreover, MIE technique does not compromise on the light emitting properties of p-Si in contrary to that prepared by stain etching. p-Si prepared by LIE technique, which also is an electrodeless technique like MIE technique, gives visible photoluminescence (PL) at room temperature similar to the p-Si prepared by MIE.  Light emitting p-Si, fabricated using these two electrodeless techniques of MIE and LIE, should be studied to investigate the structural difference in resulting p-Si. This will be helpful in understanding various mechanisms involved in p-Si fabrication and will enable scientist to explore the application of p-Si beyond the traditional applications. A comparative study will also provide knowledge to control the pore size \& other parameter to tailor different properties of p-Si.

	Though the basic p-Si fabrication mechanism is the same in LIE and MIE process (dissolution of Si atoms from Si wafer as a result of reaction with HF acid), there are some similarities and dissimilarities between the two processes and can be summarized as follows. In the LIE method, n-type crystalline Si substrate, while dipped in HF acid solution, is irradiated with laser light thereby creating excess photoexcited holes on the irradiated surface. These holes initiate the etching process [29] and result in the formation of p-Si that contains Si nanostructures. On the other hand in MIE, first metal nano particles are grown on Si wafer surface before the Si wafer is immersed in etchant solution (HF + H$_2$O$_2$). Detailed mechanism of pore formation in LIE and MIE have been reported in literature [29, 33] but a comparative study of pore formation using both of these techniques is not available and is the theme of present paper.

	Aim of this paper is to present basic properties of p-Si fabricated using MIE technique and compare its morphological and PL properties with those of p-Si prepared by LIE technique earlier [29]. A comparison of pore size is presented using scanning electron microscopy (SEM) with explanation for the existing difference in pore size in samples prepared using the two techniques. A cross- (X-) sectional SEM, done to investigate the pore growth process shows submicron sized pore formation in MIE samples.  A uniformly distributed, well-aligned Si nano wire like structures are seen from which a possibility of visible light emission has also been investigated using PL spectroscopy.

\section{Experimental details}
\ \ \ \ p-Si samples have been prepared using above-mentioned two techniques namely LIE and MIE. The LIE samples were prepared from commercially available wafers of n- Si(100) having resistivity of 3–5 $\Omega$ cm. These wafers were cleaned in acetone and ethanol to remove impurities prior to starting the porosification process. Mounted on two teflon plates, these wafers were immersed in the HF acid (40 \%) in a plastic container. The LIE [27, 29] was performed by using an argon ion laser beam (photon energy of 2.41 eV) focused onto the Si wafer with laser power density of 1.76 kW/cm$^2$. Two samples were prepared by irradiating the wafers for 45 minutes (\emph{sample: L45}) and 60 minutes (\emph{sample: L60}) with the abovementioned etching laser power density.

The MIE samples were prepared by silver (Ag) -assisted chemical etching of n-Si (100) wafer. The cleaned wafers (as discussed above) were immersed in HF solution to remove thin oxide layer formed at surface. These wafers were dipped in solution containing 4.8 M HF\& 5 mM AgNO$_3$ for one minute at room temperature to deposit Ag nano particles (AgNPs). The AgNPs deposited samples were then kept for etching in an etching solution containing 4.6 M HF \& 0.5 M H$_2$O$_2$ for 45 minutes (sample: M45) and 60 minutes (sample: M60). Etched wafers were transferred in HNO$_3$ acid to dissolve Ag metal. Then the samples were dipped into HF solution to remove oxide layer induced by nitric acid used in above step. Surface morphology of all the four samples have been characterized using Field-Emission Scanning Electron Microscopy (SEM), supra55 Zeiss. The PL from MIE sample was recorded with 325 nm laser excitation source using DongWoo optron 80K PL system at room temperature.

\section{Results and Discussion}

\ \ \ \ Surface morphologies, observed using SEM, of all the four samples (L45, L60, M45 \& M60) have been shown in Fig. 1.  Figure 1a and 1b shows SEM images of samples prepared using LIE technique whereas the SEM images of MIE samples are shown in Fig. 1c and 1d. The scale bars in Fig. 1a – 1d correspond to 10 $\mu$m. Porous Si with pore size of $\sim$10 $\mu$m can be seen in Fig. 1a and 1b corresponding to samples L45 \& L60 respectively. Comparatively wider and connected pores [29] can be seen in Fig. 1b as compared to Fig. 1a due to increased etching time. For a given laser power density, when the etching time is increased, more Si atoms gets dissolved in HF solution due to chemical etching reaction to result in deeper \& wider pores. It is important here to mention that in the absence of laser, etching does not take place and porosification is not possible. An SEM image of such a control sample is shown in Fig. S1 in the SM. Effect of etching time and laser power density on LIE process have been studied in detail and has been reported earlier [27, 29]. On the other hand, Figs.1c and 1d show SEM images corresponding to MIE samples for etching times of 45 minutes (sample M45) and 60 minutes (sample M60) respectively. Submicron sized pores are formed in MIE samples in contrast with wider pores for the case of LIE samples (Fig. 1a-1b). Wider pores are formed during LIE because porosofication can take place in lateral as well as vertical direction whereas etching lead by AgNPs during MIE take place faster in vertically downward direction rather than laterally to result in relatively smaller pores in MIE.

\begin{figure}[h]
\begin{center}
\includegraphics[height=10.5cm]{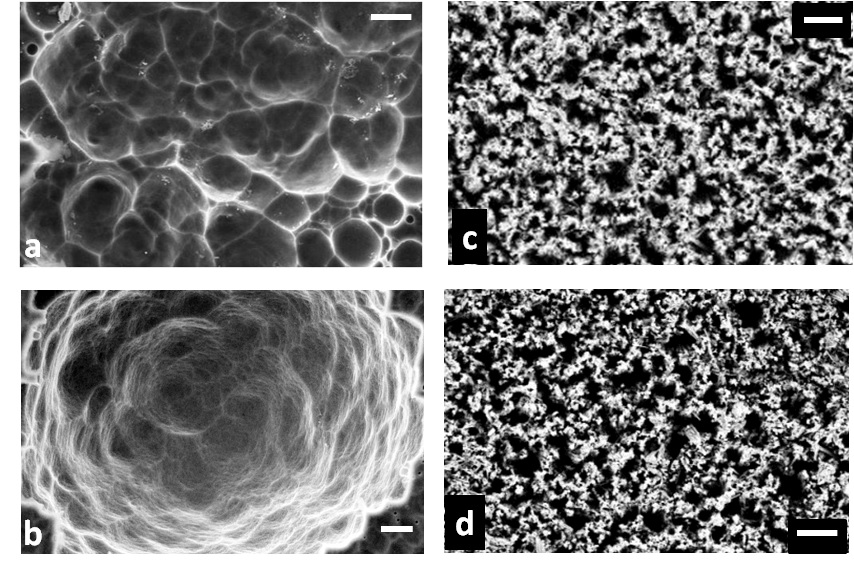}
\caption{SEM images from  (a) LIE sample with etching time 45 minutes, (b) LIE sample with etching time 60 minutes, (c) MIE sample with etching time 45 minutes  and (d) MIE sample with etching time 60 minutes. All the scale bars correspond to 10 $\mu$m}
\label{} 
\end{center}
\end{figure}
Comparison of SEM images in Fig. 1 corresponding to LIE samples (Fig. 1a-1b) and MIE samples (fig1c-1d) also reveals that the pores are densely packed for MIE samples as compared to LIE samples. It is also evident that the pores are relatively uniform in size for MIE samples whereas the pores formed using LIE is not uniform. This difference can be understood as follows. As known, the LIE process is a laser power dependent process [29]. A laser beam having Gaussian intensity profile is used for LIE of Si wafer in HF acid. The Gaussian laser beam with maximum intensity at the center of the beam induces highest etching rate at the center as compared to that near the edge of the laser beam. As a result, a variation in the pores’ size is expected in LIE samples. However this is not very clear from the SEM images shown in Fig. 1a-1b, a distribution in size of Si nanostructures contained in LIE samples has been reported to explain observed Raman line-shape earlier [39–42]. On the other hand, the MIE is initiated by AgNPs on the Si wafer deposited prior to chemical etching in etchant solution as discussed in the experimental section. It can be understood that the pore size and pore density in the case of MIE samples are direct manifestation of AgNPs’ size and distribution of AgNPs on the wafer. Furthermore, LIE technique is suitable for localized growth of p-Si and MIE technique is suitable if one needs p-Si over a relatively large area. Size of samples prepared using LIE is controlled by the spot size of laser and are typically are $\sim$100 $\mu$m in size (sample size) due to which it is not possible to obtain a X-sectional image of LIE samples. On the other hand, MIE samples give us the option to investigate the pores’ growth by analyzing the X-sectional images.

\begin{figure}[h]
\begin{center}
\includegraphics[height=10.5cm]{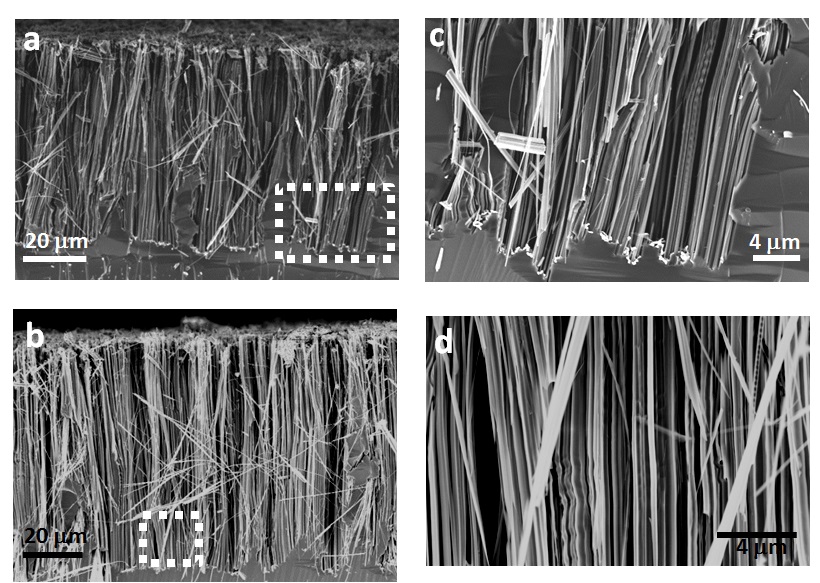}
\caption{Cross-sectional EM images from  (a) MIE sample with etching time 45 minutes, (b) MIE sample with etching time 60 minutes, (c) magnified view of selection portion of (a) and (d) magnified view of selection portion of (b) }
\label{raman} 
\end{center}
\end{figure}
	X-sectional SEM images of the MIE samples are shown in Fig. 2a-2b corresponding to samples M45 and M60 respectively. A very clear nanowire (submicron sized) like features can be seen for both the samples. The scale bars on Figs. 2a-2b are of 20 $\mu$m and those on Figs. 2c-2d are of 4 $\mu$m in length. Approximate lengths of the nanowires in for sample M45 is 60 $\mu$m whereas that for M60 it is 80 $\mu$m. This increase in nanowires’ (or pore) length is a direct manifestation of longer metal induced chemical etching of Si wafer. The parallel pores in sample M60 are more uniform as compared to sample M45. To have a closer look at the nanowires formed, the X-sectional images were magnified as shown in Fig. 2c-2d. Figures 2c-2d show the magnified view of the portion marked in figs 2a-2b respectively. It is clear from Fig. 2c-2d also that aligned uniform nanowires are formed in sample M60 \& sample M45. Such kind of parallel Si nanowires may be used as field emitters [43] for various applications. As discussed above, the nanowires are formed parallel to each other when prepared using MIE method because the growth of such nanowires are initiated and governed by the distribution and size of AgNPs grown prior to chemical etching. The X-sectional images for LIE samples cannot be presented here due to sample constraints because it is not possible to prepare the LIE samples for X-sectional SEM imaging.

	The SEM results discussed in Fig. 2, indicates that the dimensions of Si nanowires prepared as a result of MIE might be sufficiently smaller to show quantum confinement effect. PL measurement was carried out to investigate the light emitting properties of samples prepared by MIE technique. It is important here to mention that samples prepared using LIE also show visible PL as reported earlier [26, 38, 39, 42]. A visible PL at room temperature form semiconductor nanostructures is used as a signature of quantum confinement effect. Figure 3 shows the room temperature PL spectrum of the MIE sample M45. Figure 3 shows a PL peak centered at 1.96 eV and is $\sim$ 370 meV broad. The PL peak in visible region with 370 meV of broadening is a clear indication of confinement effect in the MIE sample. Inset of Fig. 3 shows the red color emission corresponding to 1.96 eV PL emission under illumination of 325 nm laser. The image shown in inset has been captured during the PL recording with UV laser (325 nm excitation). Similar kind of visible PL has been observed from Si nanostructures prepared using LIE technique as reported earlier [39]. Figure 3 shows only one peak which is in contrast with a two peak behavior for LIE samples as reported earlier [39]. 
\begin{figure}
\begin{center}
\includegraphics[height=8.0cm]{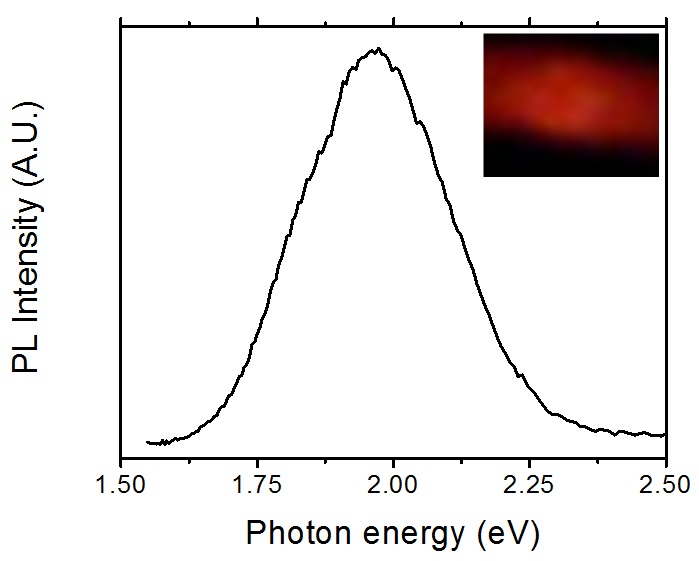}
\caption{Photoluminescence spectrum from MIE sample with etching time of 45 minutes (M45). Inset shows the photographic image of sample M45 under 325 nm laser illumination.}
\label{tem} 
\end{center}
\end{figure}The number of PL peaks in a spectrum provides information about the available sizes in the nanostructures’ sample. Single PL peak in MIE sample as shown in Fig. 3 means that the Si nanostructures are uniform and does not contain a range of sizes. In contrast, it is reported [39] that a double PL peak is observed from LIE samples due to two different dominant nanostructure sizes available in LIE samples. LIE samples contain different sized Si nanostructures because a Gaussian beam is used to prepare the sample using LIE. Variation in etching laser power density in a Gaussian beam results in Si nanostructures of different sizes in different region of p-Si sample. On the other hand, the nanostructures’ sizes are controlled by metal nanoparticles for MIE samples and result in uniform Si nanostructure.

\section{Conclusions}
  
In summary, a comparison of p-Si prepared using LIE and MIE techniques reveals that smaller and uniform pores are formed in MIE samples as compared to LIE ones. During the p-Si formation using MIE technique, the pores start forming from the sites (on wafer) where silver nanoparticles are present. Uniformly distributed silver nanoparticles lead the etching process and pores proceed mainly in the downward direction making etching more favorable in the vertical direction and lesser etching take place in the lateral direction (parallel to the wafer surface). On the other hand, in LIE method the etching take place in both the direction (vertically down the wafer and parallel to the wafer), of course with different etching rates. As a result in the same time the pores formed in LIE samples are wider as compared to those formed on MIE samples. In addition, the etching initiates at silver nanoparticles sites distributes on the Si surface in the case of MIE samples to result in relatively uniform pore distribution as compared to LIE samples. A laser beam having Gaussian profile of intensity used for LIE gives rise a distribution in pore size in LIE samples \& Si nanostructures are fabricated. A X-sectional SEM image from MIE samples show that well aligned Si nanowires are formed with increasing sizes when etching time is increased. These parallel Si nanowires may be used for field emission devices and opens scope for further studies.  The Si nanowires in the p-Si samples prepared by MIE show visible PL at 1.96 eV (red emission) at room temperature possibly due to weak quantum confinement effect.  The PL spectrum show only single peak for MIE samples in contrary with already reported double PL peak for LIE samples giving confirmation that the Si nanowires are of relatively uniform size in MIE samples as compared to LIE samples.
\section*{Acknowledgements}
 Authors are thankful to Sophisticated Instrument centre (SIC), IIT Indore (IIT Indore) for providing Photoluminescence and FESEM measurement facilities. Authors acknowledge technical support from Mr. Farhan Babu (IIT Indore) in SEM experiment.

\end{document}